\documentclass[draft,published,notoc]{JHEP} 

\usepackage{epsfig}

\newcommand\fverb{\setbox\pippobox=\hbox\bgroup\verb}
\newcommand\fverbdo{\egroup\medskip\noindent%
			\fbox{\unhbox\pippobox}\ }
\newcommand\fverbit{\egroup\item[\fbox{\unhbox\pippobox}]}
\newbox\pippobox

\title{Confining strings and RG flow in the 
(2+1)-dimensional Georgi-Glashow model
and its $SU(3)$-generalization}

\author{N. Agasian\thanks{
Partially supported by RFBR grant 00-02-17836 and INTAS grant Open Call 2000,
project No. 110.}\\
State Research Center Institute of Theoretical 
and Experimental Physics,\\ 
B. Cheremushkinskaya 25, RU-117 218 Moscow, Russia\\
E-mail: \email{agasyan@heron.itep.ru}}

\author{D. Antonov\thanks{
Permanent address: ITEP, B. Cheremushkinskaya 25, 
RU-117 218 Moscow, Russia. 
Partially supported by INTAS grant Open Call 2000,
project No. 110.}\\
INFN-Sezione di Pisa, Universita degli studi di Pisa,\\
Dipartimento di Fisica, Via Buonarroti, 2 - Ed. B - I-56127 Pisa, Italy\\
E-mail: \email{antonov@df.unipi.it}} 

\received{\today} 		
\accepted{\today}		

\preprint{JHEP/022A/0501}	
\abstract{Confining strings and RG flow at finite temperature 
are investigated in the (2+1)-dimensional 
Georgi-Glashow model. This is done
in the limit when the electric coupling constant is much larger 
than the square root of mass of the Higgs field, but much smaller
than the vacuum expectation value of this field.  
The modification of the Debye mass of the dual photon 
with respect to the case when it is considered to be negligibly small
compared to the Higgs mass, is found. Analogous modifications 
of the potential of monopole 
densities and string coupling constants are found.
At finite temperature, the mass of the Higgs field scales according 
to a novel RG equation. It is checked that in the limit
when the original theory is reduced by the RG flow to the 2D $XY$ model,
the so-evolved Higgs mass is still much smaller than the 
squared electric coupling constant. The $SU(3)$-theory of confining strings 
is also discussed within the same approximations.}


\begin{document}
\maketitle

\section{Introduction}
(2+1)-dimensional Georgi-Glashow model is known to possess confining 
properties at any values of electric coupling 
due to the presence of magnetic monopoles~\cite{pol}. 
In the limit of infinitely heavy Higgs field, 
string representation of the Wilson loop in this model is 
therefore the same as in compact 
QED, which has been constructed in Ref.~\cite{cs}. One of the aims of the 
present paper (which will be realized in Section 2) 
is to generalize the results of Ref.~\cite{cs} to the 
case when the mass of the Higgs field is large, compared to the Debye
mass of the dual photon, but nevertheless finite. In this case,
the respective dual field theory describing the grand canonical 
ensemble of monopoles has been obtained in Ref.~\cite{dietz}, and we 
shall use it as a starting point in our analysis. Another aim of the 
present paper (which will be realized in Section 3) 
is to investigate the critical properties of the model under 
study at finite temperature by means of the RG flow equations. 
{\it A priori}, 
it is clear that the mass of the Higgs field should obey its own 
RG flow equation, which will be derived and analysed. After that, 
in Section 4, the obtained 
results, regarding confining strings at the finite Higgs mass, 
will be generalized to the case of the $SU(3)$ Georgi-Glashow model.
Finally, the main results of the paper will be summarized in the 
Conclusions.

\section{Monopole potential and confining strings}

The Euclidean action of the (2+1)-dimensional Georgi-Glashow 
model has the form \cite{pol}

\begin{equation}
\label{GG}
S=\int d^3x\left[\frac{1}{4g^2}\left(F_{\mu\nu}^a\right)^2+
\frac12\left(D_\mu\Phi^a\right)^2+\frac{\lambda}{4}\left(
\left(\Phi^a\right)^2-\eta^2\right)^2\right],
\end{equation}
where the Higgs field $\Phi^a$ transforms by the adjoint representation, 
and $D_\mu\Phi^a\equiv\partial_\mu\Phi^a+\varepsilon^{abc}A_\mu^b
\Phi^c$.
In the one-loop approximation, the 
partition function of this theory reads~\cite{dietz}

$$
{\cal Z}=1+\sum\limits_{N=1}^{\infty}\frac{\zeta^N}{N!}
\left[
\prod\limits_{i=1}^{N}\int d^3z_i\sum\limits_{q_i=\pm 1}^{}\right]
\times
$$

\begin{equation}
\label{1}
\times\exp\left\{-\frac{g_m^2}{2}\left[\int d^3xd^3y\rho_{\rm gas}({\bf x})
D_0({\bf x}-{\bf y})
\rho_{\rm gas}({\bf y})-
\sum\limits_{{a,b=1\atop a\ne b}}^{N}
D_m({\bf z}_a-{\bf z}_b)
\right]\right\}.
\end{equation}
Here, $g_m$ is the magnetic coupling constant of dimensionality 
$[{\rm length}]^{1/2}$ related to the electric one $g$
according to the equation $gg_m=4\pi$, $\rho_{\rm gas}({\bf x})=
\sum\limits_{a=1}^{N}q_a\delta\left({\bf x}-{\bf z}_a\right)$
is the density of monopole gas with $q_a$'s standing for the 
monopole charges in the units of $g_m$. Next, in Eq.~(\ref{1}), 
$m=\eta\sqrt{2\lambda}$ is the mass of the Higgs boson and 

\begin{equation}
\label{zeta}
\zeta=\frac{m_W^{7/2}}{g}\delta\left(\frac{\lambda}{g^2}\right)
{\rm e}^{-(4\pi/g^2)m_W\epsilon\left(\lambda/g^2\right)}
\end{equation}
is the statistical weight of a single monopole (else called fugacity)
with $m_W=g\eta$ being the mass of the $W$-boson.
Here, $\epsilon$ is a slowly varying function equal to unity at the 
origin [{\it i.e.} in the Bogomolny-Prasad-Sommerfield 
(BPS) limit]~\cite{bps} and $1.787\ldots$ at infinity~\cite{kirk}, whereas 
the function $\delta$ is determined by the loop corrections.
Finally, in Eq.~(\ref{1}), $D_0({\bf x})\equiv1/(4\pi|{\bf x}|)$ 
is the Coulomb propagator, and 
$D_m({\bf x})\equiv{\rm e}^{-m|{\bf x}|}/(4\pi|{\bf x}|)$
is the propagator of the Higgs boson.

The effective field theory describing the grand canonical 
partition function~(\ref{1}) can easily be obtained and reads~\cite{dietz}

\begin{equation}
\label{2}
{\cal Z}=\int {\cal D}\chi{\cal D}\psi\exp\left\{-\int d^3x\left[
\frac12(\nabla\chi)^2+\frac12(\nabla\psi)^2+\frac{m^2}{2}\psi^2-
2\zeta{\rm e}^{g_m\psi}\cos(g_m\chi)\right]\right\},
\end{equation}
where $\chi$ is the dual photon field, whereas the field $\psi$ is an
additional one. The latter field can be integrated out in the 
limit 

\begin{equation}
\label{in1}
g\gg\sqrt{m}
\end{equation} 
when the exponent in the last term 
on the R.H.S. of Eq.~(\ref{2})
can be approximated by the terms not higher than the linear one.
Let us clarify the latter statement. It is based on an obvious
argument that the configuration of the $\psi$-field dominating 
in the partition function~(\ref{2}) is the one, at which 
every term of the action is of the order of unity. Then, demanding this 
for the kinetic term, $\int d^3x (\nabla\psi)^2\sim 1$, we see that the 
characteristic wavelength $L$ of the field $\psi$ is related to the 
amplitude of this field as $L\sim |\psi|^{-2}$. Substituting this
estimate into the second condition $\int d^3x m^2\psi^2\sim
L^3 m^2|\psi|^2\sim 1$, we get $|\psi|\sim m^{1/2}$. The exponent
of the $\psi$-field can be treated in the linear approximation 
when $g_m |\psi|\ll 1$, which upon the substitution of the 
obtained estimate for $|\psi|$ yields the above criterion~(\ref{in1}).

In such a limit, 
Gaussian integration over the field $\psi$ yields~\footnote{From
now on, we omit the inessential normalization factors, implying that
those are included in the respective integration measures.}

$$
{\cal Z}=\int {\cal D}\chi\exp\left\{-\int d^3x\left[
\frac12(\nabla\chi)^2-
2\zeta\cos(g_m\chi)\right]+\right.
$$

\begin{equation}
\label{3}
\left.+2(g_m\zeta)^2\int d^3xd^3y\cos(g_m\chi
({\bf x}))D_m({\bf x}-{\bf y})\cos(g_m\chi({\bf y}))\right\}.
\end{equation}
Clearly, the last term here represents the correction to the standard
result~\cite{pol}. It stems from the fact that the mass of the 
Higgs field was considered to be not infinitely large compared to the
standard Debye mass of the dual photon, $m_D=g_m\sqrt{2\zeta}$.
The respective correction to $m_D$ is positive, and the square of the 
full mass reads: 

\begin{equation}
\label{M}
M^2=m_D^2\left(1+\frac{m_D^2}{m^2}\right).
\end{equation} 
Clearly, this result is valid at $m_D\ll m$ and reproduces $m_D^2$ 
in the limit $m\to\infty$. However, since we suppose now that $m$ is
not infinitely large, but is rather bounded from above by $g^2$,
it looks reasonable to seek for the leading-order 
corrections in $m_D/m$ to various quantities.

Another relation between the dimensionful parameters in the 
model~(\ref{GG}), we shall adapt for the following analysis, is 

\begin{equation}
\label{in2}
g\ll\eta.
\end{equation}
Clearly, this inequality parallels the requirement that $\eta$ should
be large enough to ensure the spontaneous symmetry breaking from $SU(2)$
to $U(1)$. In particular, from the inequalities~(\ref{in1}) and (\ref{in2}) 
we immediately obtain:

\begin{equation}
\label{lambda}
\frac{\lambda}{g^2}\sim\left(\frac{m}{m_W}\right)^2\ll\left(\frac{g}{\eta}
\right)^2\ll 1.
\end{equation}
This means that we are working in the regime of the Georgi-Glashow 
model close to the BPS limit~\cite{bps}.

Note further that in the limit~(\ref{in1}), 
the dilute gas approximation holds perfectly. 
Indeed, this approximation implies
that the mean distance between monopoles, equal to $\zeta^{-1/3}$, 
should be much larger than the inverse mass of the $W$-boson. By virtue of 
Eq.~(\ref{zeta}) and the fact that the function $\epsilon$ is of the order
of unity, we obtain that this 
requirement is equivalent to the following one: 

\begin{equation}
\label{small}
\sqrt{\frac{\eta}{g}}\delta\left(\frac{\lambda}{g^2}\right)
{\rm e}^{-4\pi\eta/g}\ll 1.
\end{equation}
It is known~\cite{ks} that at $\lambda\ll g^2$ [{\it cf.} Eq.~(\ref{lambda})],
the function $\delta$ diverges. However, it diverges in such a way 
that at $g\ll\eta$, the growth of $\delta$ does not spoil the 
exponential smallness of the L.H.S. of Eq.~(\ref{small})~\cite{commun}.
Another consequence of this fact is that in the limit~(\ref{in1}), (\ref{in2})
of the Georgi-Glashow model, the Debye mass 
of the dual photon, $m_D$, remains exponentially small.
In particular, the inequality $m_D\ll m$, under which the full mass~(\ref{M})
was derived, holds due to this smallness. Also, due to the same reason,
the mean field approximation, under which the effective field
theory~(\ref{2}) is applicable, remains valid as well with the 
exponential accuracy. Indeed, 
this approximation implies that in the Debye 
volume $m_D^{-3}$~\footnote{In this discussion, the difference
between $m_D$ and $M$ is unimportant.} there should contain 
many particles~\cite{pol}. Since the average density of monopoles
is equal to $2\zeta$ [which can be seen either by calculating it directly
according to the formula $V^{-1}\partial\ln{\cal Z}/\partial\ln\zeta$,   
applied to Eq.~(\ref{3}) at $m_D\ll m$, or 
from the remark following after Eq.~(\ref{7}) below], 
we arrive at the requirement $\zeta m_D^{-3}\gg 1$. 
Substituting the above-obtained value for $m_D$, we see that the 
criterion of applicability of the mean field approximation 
reads $g^3\gg\sqrt{\zeta}$. Owing to the above-discussed exponential
smallness of $\zeta$, this inequality is satisfied.

Let us now derive the potential of monopole densities 
corresponding to the partition function~(\ref{3}). This can be done
by inserting into the original expression~(\ref{1}) a unity
of the form $1=\int {\cal D}\rho\delta(\rho-\rho_{\rm gas})$, 
exponentiating the $\delta$-function upon the introduction of a 
Lagrange multiplier, 
and integrating then out all the fields except of the dynamical 
monopole density $\rho$. This procedure can be seen to be equivalent 
to the substitution

$$
\exp\left[-\frac12\int d^3x(\nabla\chi)^2\right]=
$$

\begin{equation}
\label{4}
=\int {\cal D}\rho
\exp\left[-\frac{g_m^2}{2}\int d^3xd^3y\rho({\bf x})D_0({\bf x}-
{\bf y})\rho({\bf y})-ig_m\int d^3x\chi\rho\right]
\end{equation}
into Eq.~(\ref{3}) with the 
subsequent integration over the field $\chi$, which then  
plays just the r\^ole of the Lagrange multiplier.
In particular, the substitution~(\ref{4}) 
removes the kinetic term of the field
$\chi$, owing to which 
this integration can be performed in the saddle-point approximation.
The respective saddle-point 
equation should be solved iteratively in $g_m$ by setting $\chi=
\chi^{(0)}+g_m\chi^{(1)}$.
This eventually yields the following expression for the partition function
in terms of the integral over monopole densities:

\begin{equation}
\label{5}
{\cal Z}=\int {\cal D}\rho\exp\left\{-\left[
\frac{g_m^2}{2}\int d^3xd^3y\rho({\bf x})D_0({\bf x}-
{\bf y})\rho({\bf y})
+V[\rho]\right]\right\}.
\end{equation}
The monopole potential $V[\rho]$ here reads

$$
V[\rho]=\int d^3x\left[\rho{\,}{\rm arcsinh}
\varrho-
2\zeta\sqrt{1+\varrho^2}\right]-
$$

\begin{equation}
\label{6}
-2(g_m\zeta)^2
\int d^3xd^3y\sqrt{1+\varrho^2({\bf x})}
D_m({\bf x}-{\bf y})\sqrt{1+\varrho^2({\bf y})},
\end{equation}
where $\varrho\equiv\rho/(2\zeta)$.
The last term on the R.H.S. of this equation is again a leading
$(m_D/m)$-correction to the respective expression in the limit when $m$
is sent to infinity.
In the dilute gas approximation, $|\rho|\ll\zeta$, 
Eq.~(\ref{6}) becomes a simple quadratic functional:

$$V[\rho]\to\frac12\left[\frac{1}{2\zeta}-\left(\frac{g_m}{m}
\right)^2\right]\int d^3x\rho^2\simeq\frac{g_m^2}{2M^2}\int d^3x\rho^2,$$
where the last equality is implied within the leading $(m_D/m)$-approximation
adapted.
This leads to the simple expression for the generating functional
of correlators of the monopole densities within these two approximations:

$$
{\cal Z}[j]\equiv
$$

$$\equiv\int {\cal D}\rho\exp\left\{-\left[
\frac{g_m^2}{2}\int d^3xd^3y\rho({\bf x})D_0({\bf x}-
{\bf y})\rho({\bf y})
+\frac{g_m^2}{2M^2}\int d^3x\rho^2+
\int d^3xj\rho\right]\right\}=$$

\begin{equation}
\label{7}
=\exp\left[-\frac{M^2}{2g_m^2}\int d^3xd^3yj({\bf x})
j({\bf y})\partial^2 D_M({\bf x}-{\bf y})\right].
\end{equation}
[Sending for a while $m_D$ to zero (since it is exponentially small), 
we get from Eq.~(\ref{7}): 
$\left<\rho({\bf x})\rho(0)\right>=2\zeta\delta({\bf x})$. 
This means that with the exponential accuracy 
the average density of monopoles is equal to $2\zeta$, which can also be 
seen directly from the $(|\rho|\ll\zeta)$-limit of Eq.~(\ref{6}).]
In particular, the Wilson loop reads:

$$
\left<W({\cal C})\right>=\left<W({\cal C})\right>_{\rm free}{\cal Z}[i\eta]=
$$

\begin{equation}
\label{8}
=\exp\left\{-\frac{g^2}{2}\left[\frac{M^2}{2}\int\limits_{\Sigma}^{} 
d\sigma_{\mu\nu}({\bf x})
\int\limits_{\Sigma}^{} d\sigma_{\mu\nu}({\bf y})+\oint\limits_{{\cal C}}^{}
dx_\mu \oint\limits_{{\cal C}}^{}dy_\mu\right]D_M({\bf x}-{\bf y})
\right\}.
\end{equation}
This equation can straightforwardly be derived by making use of the 
formula 

$$\partial_\mu\eta({\bf x})=2\pi\varepsilon_{\mu\nu\lambda}
\left[2\partial_\nu^x\oint\limits_{{\cal C}}^{}dy_\lambda
D_0({\bf x}-{\bf y})-\int\limits_{\Sigma}^{}d\sigma_{\nu\lambda}({\bf y})
\delta({\bf x}-{\bf y})\right].$$
Here, $\eta({\bf x})=2\pi\varepsilon_{\mu\nu\lambda}\partial_\mu^x
\int\limits_{\Sigma}^{}d\sigma_{\nu\lambda}({\bf y})D_0({\bf x}-{\bf y})$
is the solid angle under which an arbitrary surface $\Sigma$ 
spanned by the contour ${\cal C}$ shows up to the observer located at the 
point ${\bf x}$. Also, in Eq.~(\ref{8}), 
$\left<W({\cal C})\right>_{\rm free}=\exp\left[-\frac{g^2}{2}
\oint\limits_{{\cal C}}^{}dx_\mu \oint\limits_{{\cal C}}^{}dy_\mu 
D_0({\bf x}-{\bf y})\right]$ is the contribution of the free photons
to the Wilson loop. The explicit $\Sigma$-dependence of the R.H.S.
of Eq.~(\ref{8}) appearing in the dilute gas approximation 
becomes eliminated by the summation over branches of the 
arcsinh-function in the full monopole action~(\ref{5})-(\ref{6}).
This is the main principle of correspondence between fields
and strings in compact QED proposed in Ref.~\cite{cs} in the 
language of the Kalb-Ramond field $h_{\mu\nu}$, 
$\varepsilon_{\mu\nu\lambda}\partial_\mu h_{\nu\lambda}\propto\rho$.

As far as the string tension and the inverse coupling constant of
the rigidity term~\cite{rid} are concerned, those can be evaluated 
upon the derivative expansion of the $\Sigma$-dependent part of 
Eq.~(\ref{8}). 
By virtue of the general formulae from Ref.~\cite{aes} we obtain 

\begin{equation}
\label{sigalfa}
\sigma=4\pi g^2M~~ {\rm and}~~
\alpha^{-1}=-\frac{\pi g^2}{2M},
\end{equation}
respectively. Clearly, both of 
these quantities
represent the modifications of the standard 
ones, corresponding to the limit when $m$ is considered to be 
infinitely large {\it w.r.t.} $m_D$. The standard expressions
follow from Eq.~(\ref{sigalfa}) upon the substitution 
into this equation $M\to m_D$.

\section{RG flow at finite temperature}
At non-zero temperature, the path integral~(\ref{2}) is defined with 
periodic boundary conditions in the interval $0<t<\beta\equiv T^{-1}$.
First straightforward estimate concerns the critical temperature
of the Berezinskii-Kosterlitz-Thouless phase transition from the 
plasma phase to the phase where monopoles and antimonopoles 
are bound into molecules. Denoting by ${\sf x}$ the spatial 2D-vector
and taking into account that at $|{\sf x}|\gg\beta$, $\frac{1}{|{\bf x}|}\to
-2T\ln(\mu |{\sf x}|)$, $\frac{{\rm e}^{-m|{\bf x}|}}{|{\bf x}|}\to 
2TK_0(m|{\sf x}|)$,
where $\mu$ stands for the IR cutoff, and $K_0$ denotes the modified
Bessel function, we get for the mean squared separation between 
a monopole and an antimonopole at high temperatures:

\begin{equation}
\label{r2}
\left<r^2\right>\sim\int d^2{\sf x} |{\sf x}|^{2-\frac{8\pi T}{g^2}}
\exp\left[\frac{4\pi T}{g^2}K_0(m|{\sf x}|)\right]\sim
\int\limits_{}^{\infty} d|{\sf x}| |{\sf x}|^{3-\frac{8\pi T}{g^2}}
\exp\left[\frac{(2\pi)^{3/2}T}{g^2(m|{\sf x}|)^{1/2}}{\rm e}^{-m|{\sf x}|}
\right].
\end{equation}
Clearly, the value of the critical temperature~\cite{nk} 
$T_c=g^2/(2\pi)$
(above which the integral converges, which signals on the appearance
of the monopole-antimonopole molecules)
is not affected by the Higgs-motivated exponential factor. That is 
because at arbitrary temperature, 
the latter one tends rapidly to unity at $|{\sf x}|\to\infty$.
  
Let us now proceed with the RG analysis of the leading 
$(m_D/m)$-part of the partition 
function~(\ref{3}), which has the form

$$
{\cal Z}=
$$

\begin{equation}
\label{r3}
=\int {\cal D}\chi\exp\left\{\int d^3x\left[\frac12\chi
\left(\partial_x^2+\partial_y^2+\partial_t^2\right)\chi+2\zeta
\cos(g_m\chi)+\left(\frac{g_m\zeta}{m}\right)^2\cos(2g_m\chi)\right]
\right\}.
\end{equation}
In this way, we shall apply the techniques
elaborated out in Refs.~\cite{ruk},~\cite{sy} for the case when $m$ 
is considered to be infinitely large {\it w.r.t.} $m_D$.
First, let us split the cutoff sine-Gordon field as $\chi_\Lambda=
\chi_{\Lambda'}+h$, where the field 
$h$ includes the modes with the momenta lying in the range between 
$\Lambda'$ and $\Lambda$ and with all possible 
Matsubara frequencies. Then, integrating the $h$-field out and denoting 
$f({\bf x})\equiv 1+\left(\frac{g_m}{m}\right)^2\delta({\bf x})$, one gets

$${\cal Z}=\int {\cal D}\chi_{\Lambda'}\exp\left[-\frac12\int d^3x
\left(\nabla\chi_{\Lambda'}\right)^2\right]\exp\left\{2\zeta\int d^3x
\left<\cos\left(g_m\chi_\Lambda\right)\right>_h+2\zeta^2\int d^3xd^3y\times
\right.$$

$$\left.\times
\left[\left<\cos\left(g_m\chi_\Lambda({\bf x})\right)
\cos\left(g_m\chi_\Lambda({\bf y})\right)\right>_h f({\bf x}-{\bf y})-
\left<\cos\left(g_m\chi_\Lambda({\bf x})\right)\right>_h
\left<\cos\left(g_m\chi_\Lambda({\bf y})\right)\right>_h\right]\right\},$$
where $\left<\ldots\right>_h$ means the average {\it w.r.t.} the 
action $\frac12\int d^3x(\nabla h)^2$. By calculating the 
averages, the argument of the last exponent can be written as

$$
2\zeta A(0)\int d^3x\cos\left(g_m\chi_{\Lambda'}\right)+
(\zeta A(0))^2\times
$$

\begin{equation}
\label{9}
\times\sum\limits_{k=0,1}^{}
\int d^3xd^3y\left[A^{2(-)^k}({\bf x}-{\bf y})f({\bf x}-{\bf y})-1
\right]\cos\left[g_m\left(
\chi_{\Lambda'}({\bf x})+(-)^k\chi_{\Lambda'}({\bf y})\right)\right],
\end{equation}
where $A({\bf x})=\exp\left(-\frac{g_m^2}{2}G({\bf x})\right)$
with $G({\bf x})$ denoting the propagator of the field 
$h$. [Clearly, $G$ is different from $D_0$, since the 
momenta of the field $h$ vary only 
in the finite interval $(\Lambda',\Lambda)$.] Similarly to 
Ref.~\cite{sy}, we shall investigate the generalized model where
$G(0)$ is considered to be finite. Next, 
one can pass to the integration over center-of-mass and relative-distance
coordinates, ${\bf z}=({\bf x}+{\bf y})/2$ and ${\bf u}={\bf x}-{\bf y}$,
and taking into account that 
$A^{\pm 2}({\bf u})\to 1$ at $|{\bf u}|\to\infty$, Taylor expand
in ${\bf u}$ the last integrand on the R.H.S. of Eq.~(\ref{9}). 
This yields for the respective integral the following expression:

$$
\int d^3z\left[\chi_{\Lambda'}({\bf z})\left(
\sum\limits_{\mu}^{}\eta_\mu^2\partial_\mu^2\right)
\chi_{\Lambda'}({\bf z})
+\left(\frac{g_mA(0)}{m}\right)^2
\cos\left(2g_m\chi_{\Lambda'}({\bf z})\right)\right],$$
where $\eta_\mu=\frac{g_m^2}{2}\int d^3uu_\mu^2[A^{-2}({\bf u})
f({\bf u})-1]$. 
Introducing the notation 
$\zeta_\mu^{-2}=1+2\zeta^2A^2(0)\eta_\mu$,
rescaling the field $\chi_{\Lambda'}$ 
as $\chi=\zeta_x^{-1}\chi_{\Lambda'}$ and 
momenta as ${\bf p}'=(\Lambda/\Lambda'){\bf p}$, and denoting $\gamma=\left(
\frac{\Lambda}{\Lambda'}\frac{\zeta_x}{\zeta_t}\right)^2$,
we eventually get the following 
expression for the partition function~(\ref{r3}):

$$
{\cal Z}=\int {\cal D}\chi\exp\left\{\int d^3x\left[
\frac12\chi\left(\partial_x^2+\partial_y^2+\gamma\partial_t^2\right)\chi
+\right.\right.$$

\begin{equation}
\label{10}
\left.\left.
+2\zeta A(0)\cos(g_m\zeta_x\chi)+\left(\frac{g_m\zeta A^2(0)}{m}\right)^2
\cos(2g_m\zeta_x\chi)\right]\right\}.
\end{equation}

To derive the RG equations, one should consider Eq.~(\ref{10}) at
the infinitesimal transformation $\Lambda'=\Lambda-\delta\Lambda$ and compare 
it with the original expression~(\ref{r3}).
Clearly, this comparison leads to the RG equation describing the evolution 
of $m$, which is new {\it w.r.t.} the standard equations describing 
the evolution of $\zeta$ and $g_m$. 
In this way, the following formulae, which can be straightforwardly
derived, are useful:

$$
\frac{d\zeta_x}{d\Lambda}=
\frac{g_m^4A^2(0){\cal J}_2\zeta^2}{4\pi\Lambda^5},~~
\frac{dA(0)}{d\Lambda}=\frac{\pi x}{\Lambda}\frac{\tau}{2}
\coth\frac{\tau}{2}.$$
Here, $d\Lambda\equiv-\delta\Lambda$, $x=\frac{Tg_m^2}{4\pi^2}$, 
$\tau=\frac{\Lambda}{T\sqrt{\gamma}}$, and (within the notations 
of Ref.~\cite{sy})
${\cal J}_2\equiv\pi\int\limits_{0}^{\infty}d\xi\xi^3
J_0(\xi)$
with $J_0$ standing for the Bessel function. Following Ref.~\cite{sy},
one can 
introduce one more dimensionless parameter 
$z=(2\pi)^3{\cal J}_2A^2(0)\frac{\zeta}{\Lambda^2T}$
and by denoting $t=-\ln(\Lambda/T)$
get the following system of RG equations:

\begin{equation}
\label{11}
dx=-x^3z^2dt,~~ 
dz^2=-2z^2\left(\pi x\frac{\tau}{2}\coth\frac{\tau}{2}-2\right)dt,~~
d\ln\tau=-2dt.
\end{equation}
Next, we can also introduce a novel, $m$-dependent, 
dimensionless parameter 

$$u\equiv\Lambda^{-3/4}
\sqrt{g_m\zeta m^{-1}}$$ 
and comparing Eqs.~(\ref{10}) and~(\ref{r3}) derive for it the following 
RG equation:

\begin{equation}
\label{RGu}
d\ln u=\left(\frac34-\pi x\frac{\tau}{2}\coth\frac{\tau}{2}\right)dt.
\end{equation}
This is just the announced new 
equation which describes indirectly the evolution of $m$.

In the limit $t\to\infty$
(or $\tau\to 0$), the RG flow~(\ref{11}) 
in the model under study becomes that of the 2D $XY$ model.
In particular, the BKT transition point of the $XY$ model, $z_c=0$, 
$x_c=2/\pi$, can straightforwardly be read off from Eq.~(\ref{11}). 
It is worth noting that the
critical value $x_c$ corresponds to the critical temperature 
$T_c=g^2/(2\pi)$, which was obtained above heuristically, without any
RG analysis. In the vicinity of the BKT transition point, the 
RG trajectories stemming from the integration of the first two of
Eqs.~(\ref{11}) are typical hyperbolae of the $XY$-model. They are 
defined by the equation $(x-x_c)^2-(2/\pi)^3z^2=C_1$, where $C_1$ stands for
some constant.

In the same $XY$-model limit, 
the flow of $u$ takes the form 

\begin{equation}
\label{eqforu}
d\ln u=\left(\frac34-\pi x\right)dt.
\end{equation}
This equation yields the flow of $m$ itself:
$m=Cg_m\zeta T^{-\frac32}\left(\frac{T}{\Lambda}
\right)^{\frac{Tg_m^2}{2\pi}}$, where $C$ is $g_m$-independent 
constant of integration.
The limit~(\ref{in1}) is then satisfied 
provided that the following inequality holds: $g_m^3
\left(\frac{T}{\Lambda}\right)^{\frac{Tg_m^2}{2\pi}}\ll
T^{\frac32}/(C\zeta)$. Owing to Eq.~(\ref{zeta}), it can be rewritten as 
$g_m\eta^7\left(\frac{T}{\Lambda}\right)^{\frac{Tg_m^2}{\pi}}
\ll C^{-2}T^3$. We see that in the limit of small $\Lambda$ under discussion, 
this inequality is indeed satisfied for any given $C$ 
at sufficiently small $g_m$. 
This means that in the $XY$-model limit, the RG flow 
does not lead us outside the scope of the original 
approximation~(\ref{in1}). 
Finally, in the vicinity of the BKT transition point, 
Eq.~(\ref{eqforu}) can be integrated together 
with the first two of Eqs.~(\ref{11}). This yields $u=\exp\left[
-\frac{5(x-x_c)}{4C_1}+C_2\right]$, where $C_2$ is a
certain constant independent of $C_1$.

\section{String representation of the Wilson loop in the $SU(3)$-case}
In the $SU(3)$-generalization of the Georgi-Glashow model, 
the monopole density $\rho_{\rm gas}$ entering Eq.~(\ref{1}) 
should be replaced by the following one~\cite{dw},~\cite{sn}:
$\vec\rho_{\rm gas}({\bf x})=\sum\limits_{a}^{}
{\vec q}_{i_a}\delta({\bf x}-{\bf z}_a)$.  
Here, $\vec q_i$'s are 
the root vectors of the group $SU(3)$:

$$\vec q_1=\left(\frac12,\frac{\sqrt{3}}{2}\right),~ 
\vec q_2=(-1,0),~
\vec q_3=\left(\frac12,-\frac{\sqrt{3}}{2}\right),~
\vec q_{-i}=-\vec q_i.$$
The partition function~(\ref{2}) becomes modified to

$$
{\cal Z}=\int {\cal D}\vec\chi{\cal D}\psi\times$$

$$\times\exp\left\{-\int d^3x\left[
\frac12(\nabla\vec\chi)^2+\frac12(\nabla\psi)^2+\frac{m^2}{2}\psi^2-
2\zeta{\rm e}^{g_m\psi}\sum\limits_{i=1}^{3}
\cos\left(g_m\vec q_i\vec\chi\right)\right]\right\},
$$
where $\vec\chi=\left(\chi^1,\chi^2\right)$ and 
analogously to Ref.~\cite{dw} we have assumed that the $W$-bosons
corresponding to different root vectors have the same masses.

Similarly to Section 2, 
in the limit~(\ref{in1}), the field $\psi$ can be integrated out, 
and we get the following $SU(3)$-version of Eq.~(\ref{3}):

$$
{\cal Z}=\int {\cal D}\vec\chi\exp\left\{-\int d^3x\left[
\frac12(\nabla\vec\chi)^2-2\zeta\sum\limits_{i=1}^{3}
\cos\left(g_m\vec q_i\vec\chi\right)\right]+\right.$$

\begin{equation}
\label{13}
\left.+2(g_m\zeta)^2\sum\limits_{i,j=1}^{3}\int d^3xd^3y
\cos\left(g_m\vec q_i\vec\chi({\bf x})\right)D_m({\bf x}-{\bf y})
\cos\left(g_m\vec q_j\vec\chi({\bf y})\right)\right\}.
\end{equation}
The full Debye mass stemming from the expansion of cosines 
in this equation can straightforwardly be obtained by noting that 
$\sum\limits_{i=1}^{3}(\vec q_i\vec\chi)^2=\frac32\vec\chi{\,}^2$.
It reads $M^2=m_D^2\left(1+2\frac{m_D^2}{m^2}\right)$, where 
$m_D^2=3g_m^2\zeta$ is the squared standard Debye mass in the 
limit when $m$ is sent to infinity.

In order to derive the potential of monopole densities, one should 
again perform the substitution~(\ref{4}) with the replacements 
$\chi\to\vec\chi$, $\rho\to\vec\rho$. The resulting saddle-point
equation,

$$\sum\limits_{i=1}^{3}\vec q_i\sin(g_m\vec q_i\vec\chi({\bf x}))
\left[1+2g_m^2\zeta\sum\limits_{j=1}^{3}\int d^3yD_m({\bf x}-{\bf y})
\cos(g_m\vec q_j\vec\chi({\bf y}))\right]=-\frac{i}{2\zeta}
\vec\rho({\bf x}),$$
should again be solved iteratively by setting $\vec\chi=\vec\chi^{(0)}+
g_m\vec\chi^{(1)}$. This equation can be solved {\it w.r.t.} $\vec q_i
\vec\chi$ by representing $\vec\rho$ as~\cite{epl} $\sum\limits_{i=1}^{3}
\vec q_i\rho_i$, where $\rho_1\equiv
\left(\rho^1/\sqrt{3}+\rho^2\right)/\sqrt{3}$, 
$\rho_2\equiv -2\rho^1/3$, $\rho_3\equiv
\left(\rho^1/\sqrt{3}-\rho^2\right)/\sqrt{3}$. The so-obtained solution
has the form

$$\vec q_i\vec\chi^{(0)}({\bf x})=-\frac{i}{g_m}{\,}{\rm arcsinh}{\,}
\varrho_i({\bf x}),$$

$$\vec q_i\vec\chi^{(1)}({\bf x})=\frac{i\rho_i({\bf x})}
{\sqrt{1+\varrho_i^2({\bf x})}}
\sum\limits_{j=1}^{3}\int d^3yD_m({\bf x}-{\bf y})
\sqrt{1+\varrho_j^2({\bf y})},$$
where $\varrho_i\equiv\rho_i/(2\zeta)$. This yields the desired 
representation of the partition function in the form of Eq.~(\ref{5})
with $\rho$ replaced by $\vec\rho$ and $V[\rho]$ replaced by

$$
V[\vec\rho{\,}]=\int d^3x\sum\limits_{i=1}^{3}
\left[\rho_i{\,}{\rm arcsinh}\varrho_i-
2\zeta\sqrt{1+\varrho_i^2}\right]-
$$

$$
-2(g_m\zeta)^2
\int d^3xd^3y\sum\limits_{i,j=1}^{3}\sqrt{1+\varrho_i^2({\bf x})}
D_m({\bf x}-{\bf y})\sqrt{1+\varrho_j^2({\bf y})}.
$$
Similarly to the $SU(2)$-case, in the dulite gas approximation,
$|\vec\rho{\,}|\ll\zeta$, the monopole potential becomes a quadratic
functional:

$$V[\vec\rho{\,}]\to\frac13\left(\frac{1}{2\zeta}-\frac{3g_m^2}{m^2}
\right)\int d^3x\vec\rho{\,}^2\simeq\frac{g_m^2}{2M^2}\int d^3x
\vec\rho{\,}^2,$$
where the last equality is again implied within the leading 
$(m_D/m)$-approximation.

The monopole part of the Wilson loop 
can be written as 

\begin{equation}
\label{Wloopmon}
\left<W({\cal C})\right>_{\rm mon}=\frac13\sum\limits_{\alpha=1}^{3}
\left<\exp\left(i\int d^3x\vec\rho\vec\mu_\alpha\eta\right)\right>.
\end{equation}
Here, $\vec\mu_\alpha$'s are the weight vectors of the group $SU(3)$, 
which are just 
the charges of a quark of the 
$\alpha$-th colour {\it w.r.t.} the diagonal gluons: $\vec\mu_1=\left(
-\frac12,\frac{1}{2\sqrt{3}}\right)$, $\vec\mu_2=
\left(\frac12,\frac{1}{2\sqrt{3}}\right)$, $\vec\mu_3=
\left(0,-\frac{1}{\sqrt{3}}\right)$. Note that in a derivation of 
Eq.~(\ref{Wloopmon}) we have used the identity
${\rm tr}{\,}\exp\left(i\vec a\frac{\vec\lambda}{2}\right)=
\sum\limits_{\alpha=1}^{3}\exp\left(i\vec a\vec\mu_\alpha\right)$
valid for an arbitrary vector $\vec a$.
Also, in Eq.~(\ref{Wloopmon}),    
the average is performed {\it w.r.t.} the just discussed above 
partition function expressed in terms of $\vec\rho${\,}'s.

In the dilute gas approximation, this partition function 
thus takes the form 

$$
{\cal Z}=\int {\cal D}\vec\rho\exp\left\{-\left[\frac{g_m^2}{2}
\int d^3xd^3y\vec\rho({\bf x})D_0({\bf x}-{\bf y})\vec\rho({\bf y})+
\frac{g_m^2}{2M^2}\int d^3x\vec\rho{\,}^2\right]\right\},
$$
The resulting Gaussian integration can straightforwardly be performed
and yields

$$
\left<W({\cal C})\right>_{\rm mon}
=\exp\left\{-\frac{g^2}{6}
\left[\frac{M^2}{2}\int\limits_{\Sigma}^{} 
d\sigma_{\mu\nu}({\bf x})
\int\limits_{\Sigma}^{} d\sigma_{\mu\nu}({\bf y})
D_M({\bf x}-{\bf y})+\right.\right.$$

$$\left.\left.
+\oint\limits_{{\cal C}}^{}
dx_\mu \oint\limits_{{\cal C}}^{}dy_\mu\left(D_M({\bf x}-{\bf y})-
D_0({\bf x}-{\bf y})\right)\right]
\right\},$$
where it has been taken into account that for any $\alpha$,
$\vec\mu_\alpha^2=1/3$.

An expression for the full Wilson loop can be obtained upon the 
multiplication of this result by
the contribution brought about 
by the free photons. The latter one reads   

$$\left<W({\cal C})\right>_{\rm free}=\frac13\sum\limits_{\alpha=1}^{3}
\left<\exp\left(i\vec\mu_\alpha\oint\limits_{{\cal C}}^{}dx_\mu\vec A_\mu
\right)\right>
=\exp\left(-\frac{g^2}{6}\oint\limits_{{\cal C}}^{}dx_\mu
\oint\limits_{{\cal C}}^{}dy_\mu D_0({\bf x}-{\bf y})\right),$$
where $\left<\ldots\right>$ stands for the average {\it w.r.t.} the 
action $\frac{1}{4g^2}\int d^3x\vec F_{\mu\nu}^2$.
The full Wilson loop then has 
the form of the R.H.S. of Eq.~(\ref{8}) with $g^2$
replaced by $g^2/3$. Clearly, the $SU(3)$-expressions
for $\sigma$ and $\alpha^{-1}$ can be obtained from Eq.~(\ref{sigalfa})
by performing in that equation the same replacement.

\section{Conclusions}
In the present paper, we have explored confining strings and RG flow in 
the (2+1)-dimensional Georgi-Glashow model.
This has been done in a certain limit~(\ref{in1}), (\ref{in2}) of this model 
and in the leading $(m_D/m)$-approximation, where $m_D$ is the Debye 
mass of the dual photon and $m$ is the mass of the Higgs boson. 
Both in the $SU(2)$- and $SU(3)$-cases, 
we have derived the respective Higgs-inspired corrections
to $m_D$, to the potential of monopole densities, and 
as a consequence, to the string tension and rigidity coupling constant
of the confining string. As far as the RG analysis is concerned, for the
$SU(2)$-theory there
has been obtained a novel equation describing the evolution 
of the Higgs mass. This equation takes a remarkably simple form in the 
limit when the original theory goes over to the 2D $XY$ model.
Owing to this fact, it has been checked that the evolution of 
the Higgs mass in this limit does not violate 
the adapted approximation~(\ref{in1}).

In the present paper, we have disregarded possible effects 
brought about by the $W$-bosons. Recently~\cite{Ws},~\cite{newref}, the 
influence of these bosons to the dynamics of the phase transition 
has been studied by treating them as vortices of a certain disorder
operator. In particular
for $SU(N)$-case, the respective RG-analysis has been performed  
in Ref.~\cite{newref}.
In the forthcoming paper~\cite{AA},
we plan to take into account the 
effects of $W$-bosons to the theory of confining strings at finite
temperature.

\section*{Acknowledgments}
We are greatful to A. Kovner for bringing 
Ref.~\cite{Ws} to our attention, to K. Selivanov 
for a very useful discussion, and to M. Chernodub for a valuable 
correspondence.
D.A. is indebted to A. Di Giacomo for 
many useful discussions and cordial hospitality. 
He is also greatful to 
the whole staff of the Quantum Field Theory Division
of the University of Pisa for kind hospitality and to INFN for the  
financial support.

\end{document}